\documentclass[aip,reprint]{revtex4-1}
\usepackage{graphicx}
\usepackage{color}
\usepackage{textcomp}
\draft 

\newif\ifcmnt
\cmnttrue

\ifcmnt
    \providecommand{\aucmnt}[1]{#1}
\else
    \providecommand{\aucmnt}[1]{}
\fi

\begin{document}



\title{All-silicon light-emitting diodes waveguide-integrated with superconducting single-photon detectors}

\author{Sonia Buckley}
\email[]{sonia.buckley@nist.gov}
\author{Jeffrey Chiles}
\author{Adam N. McCaughan}
\author{Galan Moody}
\author{Kevin L. Silverman}
\author{Martin J. Stevens}
\author{Richard P. Mirin}
\author{Sae Woo Nam}
\author{Jeffrey M. Shainline}
\affiliation{National Institute of Standards and Technology, Boulder, CO, USA}

\begin{abstract}
We demonstrate cryogenic, electrically-injected, waveguide-coupled Si light-emitting diodes (LEDs) operating at 1.22 $\mu$m. The active region of the LED consists of W centers implanted in the intrinsic region of a $p$-$i$-$n$ diode. The LEDs are integrated on waveguides with superconducting nanowire single-photon detectors (SNSPDs). We demonstrate the scalability of this platform with an LED coupled to eleven SNSPDs in a single integrated photonic device. Such on-chip optical links may be useful for quantum information or neuromorphic computing applications.
\end{abstract}


\maketitle 

\section{Introduction}
\label{sec:introduction}

An electrically-injected, monolithic, silicon-based light source would be of enormous benefit for optical interconnects and communications \cite{Zhou2015,Liang2010}. Despite the demand, such a source has yet to be developed due to the indirect bandgap in Si, which leads to inefficient optical transitions. For most telecommunications applications, the requirement of room-temperature operation has limited the interest in low-temperature Si-based light sources. However, there are a variety of applications for which low-temperature operation is desirable. These include quantum optics, superconducting computing \cite{Manheimer2015} and high-performance neuromorphic computing applications \cite{Shainline2017a}. These emerging applications benefit from mature semiconductor microfabrication processes for reliability and scaling. In these applications, we can take advantage of light emission processes that are only practical at cryogenic temperatures, such as emission based on defects in Si. A variety of such defects have been studied \cite{Davies1989, Shainline2007}, and electrically injected LEDs based on implanted defects and dislocations in Si have been demonstrated \cite{Bradfield1989,Libertino2000,Rotem2007,Bao2007,Lourenco2008,LoSavio2011,Sumikura2014}. In particular, the W center is a defect with a zero-phonon line at 1.22 $\mu$m, generated by implantation of Si ions \cite{Yang2010}. The W center is thought to comprise self-interstitials\cite{Carvalho2005a,Santos2016} in a trigonal geometry\cite{Davies1987}. Light-emitting diodes based on W centers implanted in the intrinsic region of a $p$-$i$-$n$ diode have previously been demonstrated \cite{Bao2007}. However, for use in photonic integrated circuits, the sources must be fabricated in a process with other active components such as detectors or modulators and must emit into a waveguide mode for routing and processing of the signal.

\begin{figure}
\includegraphics[width=6 cm]{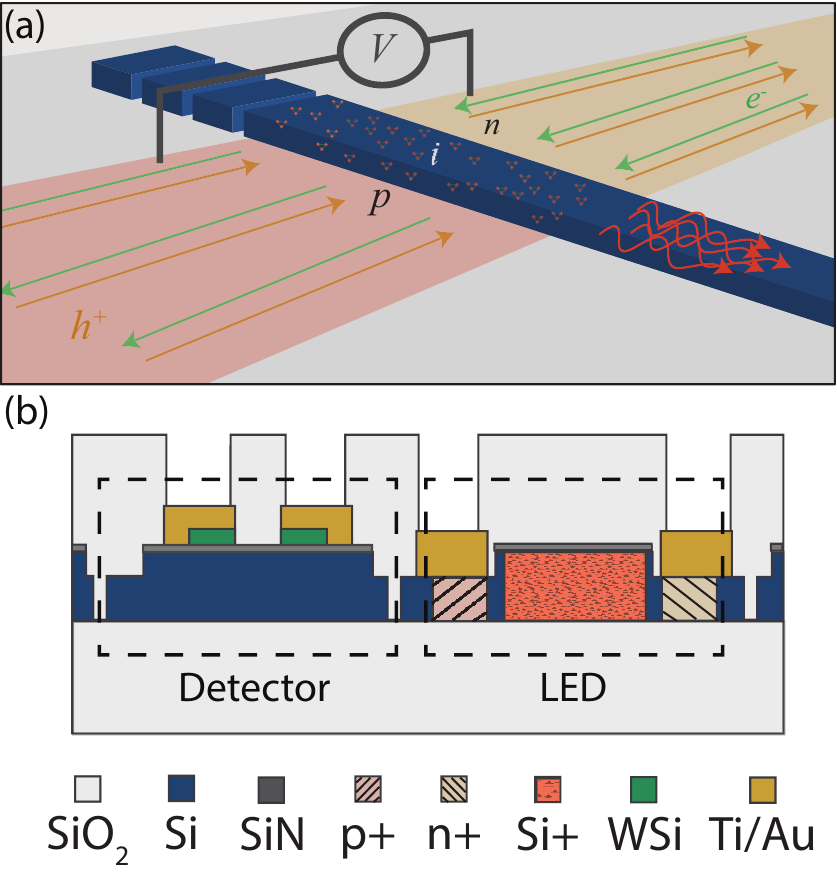}
 \caption{\label{fig:schematics}(a) Schematic of LED. (b) Schematic of process layers.}%
 \end{figure}

We demonstrate a waveguide-coupled LED based on W centers that can be used as a source for cryogenic photonic integrated circuits. A schematic showing the concept of the Si LED is shown in Fig. \ref{fig:schematics} (a). We have also integrated these LEDs with receivers consisting of waveguide-coupled superconducting nanowire single-photon detectors (SNSPDs)  \cite{Marsili2013,Shainline2017} to demonstrate a cryogenic on-chip optical link with conversion from electrical to optical and back to the electrical domain. 

\section{Fabrication}
\label{sec:fabrication}

The devices were fabricated on a 1-10 $\Omega\cdot$cm $p$-type silicon-on-insulator (SOI) wafer with a 220 nm Si layer on 2 $\mu$m buried oxide. First, alignment marks were patterned and etched in the wafer. The wafer was cleaned in buffered oxide etch (BOE) and H$_2$SO$_4$/H$_2$O$_2$ before depositing 100 nm SiO$_2$ using plasma enhanced chemical vapor deposition. This SiO$_2$ serves as a protective layer during ion implantation for the electrical and emissive dopants that comprise the LED. The areas for $p$-type dopants were patterned and implanted with B$^+$ at 168 keV and a fluence of 1.67$\times 10^{15}$ cm$^{-2}$. This implant density achieves degenerate doping, necessary for mitigation of carrier freeze-out at cryogenic temperatures. Similar patterning and implantation was carried out for the $n$-type regions (P$^+$, energy = 62 keV, fluence = 1.43 $\times 10^{15}$ cm$^{-2}$), followed by dopant activation with a rapid thermal anneal for one minute at 1100$^\circ$C and two minutes at 900$^\circ$C. With these parameters, the density of implanted dopants peaks at the surface of the waveguide shallow etch. 
The emissive center (Si$^+$, energy = 150 keV, fluence = 1.6 $\times 10^{-12}$ cm$^{-2}$) regions were then patterned and implanted. The protective oxide was removed with 6:1 BOE, and an 80 nm nitride spacer layer was deposited for electrical isolation of the nanowire detectors from the LEDs. A 3.5 nm WSi layer\cite{Baek2011} (for the SNSPDs) was then sputtered followed by a 2 nm amorphous Si protective layer. SNSPDs were patterned with 300 nm width and 100 $\mu$m length \cite{Shainline2017}. Reactive ion etching was performed using Ar and SF$_6$. Ridge waveguides were then etched through the SiN$_x$ spacer and 80 nm into the Si using a CF$_4$ chemistry. A deep etch of the Si around the structures was then performed for optical and electrical isolation. Au pads with a Ti adhesion layer were patterned for electrical contacts to both the SNSPDs and LEDs. An oxide overcladding was deposited, and vias were etched to make contact to the pads. 

Finally the W centers were annealed for 30 minutes at 250$^\circ$C \cite{Yang2010}. All lithography was performed with a 365 nm i-line stepper \cite{Shainline2017}. An overview of the process layers is shown in Fig. \ref{fig:schematics} (b). The wafer included die with four different patterns: an electrical and photoluminescence test pattern, waveguide-integrated LEDs to tapers for electroluminescence (EL) spectroscopy, waveguide-integrated LEDs to SNSPDs, and LEDs to arrays of SNSPDs for a demonstration of scalability.

\section{Experimental measurements}
 \begin{figure}

\includegraphics[width=8.5cm]{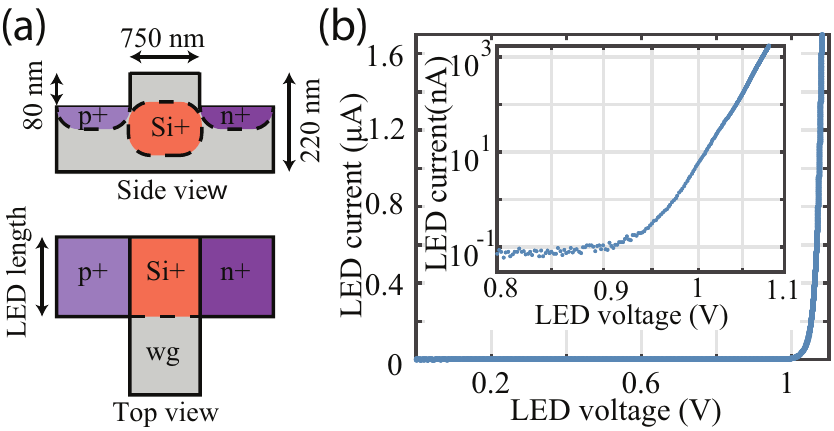}
 \caption{\label{fig:IV}(a) Schematic of an LED with the relevant dimensions indicated. (b) IV curve for an LED of length 10 $\mu$m. Inset: zoom in to the voltage range near threshold on a log $y$ scale.}%
 \end{figure}

Figure \ref{fig:IV} (a) shows a diagram of an LED, with the relevant dimensions indicated. LEDs with lengths from 0.8 $\mu$m to 100 $\mu$m were fabricated. The current-voltage characteristics of the LEDs were first tested in a sorption pump cryostat at 800 mK. The turn-on voltage was determined to be 1V. Figure \ref{fig:IV} (b) shows a typical I-V curve for an LED with 10 $\mu$m length, which was also the standard length chosen for the LEDs in the scalability experiments. The turn-on voltage (the voltage was measured for each device with the current at 1 nA) varied with the length of the device (increasing for shorter devices due to increased differential resistance). The reverse-bias leakage current was $<$ 100 pA in all measurements. 

\begin{figure}
\includegraphics[width=8.5cm]{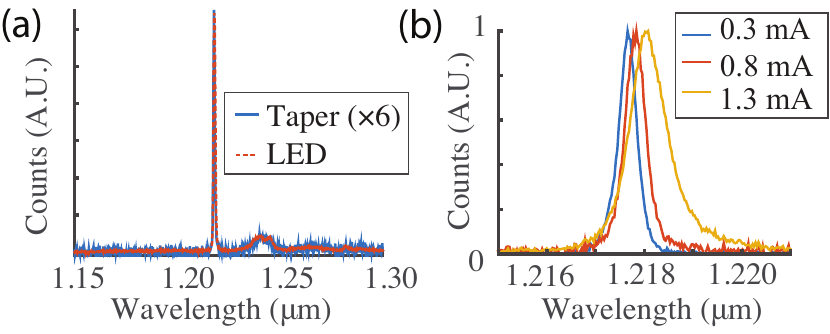}
 \caption{\label{fig:spectra}(a) EL spectrum measured above the LED (red) and above the taper (blue). (b) Higher resolution showing the linewidth of the zero phonon line of the W-center for different bias current.}%
 \end{figure}

The spectral properties of the LEDs were then investigated using a liquid-nitrogen-cooled linear InGaAs photo-diode array and spectrometer. The LED is coupled to a 1 mm-long waveguide that ends in a taper intended to scatter light. The devices were cooled in a closed cycle He cryostat at 4.2 K. The LEDs were electrically injected with a DC current, and EL was collected with a 0.6 NA objective lens. The resulting spectrum is shown in Fig. \ref{fig:spectra} (a). Light was collected from above the LED and subsequently from the taper at the end of the waveguide with six times higher intensity observed above the LED. A higher resolution spectrum of the zero phonon line is shown in Fig. \ref{fig:spectra} (b), with different bias currents normalized to have the same maximum intensity. The linewidth is 0.5 nm at 0.3 mA, broadening to 1 nm at 1.3 mA. The spectrum shifts by 0.4 nm in this range of bias currents. 

\begin{figure}
\includegraphics[width=8.5cm]{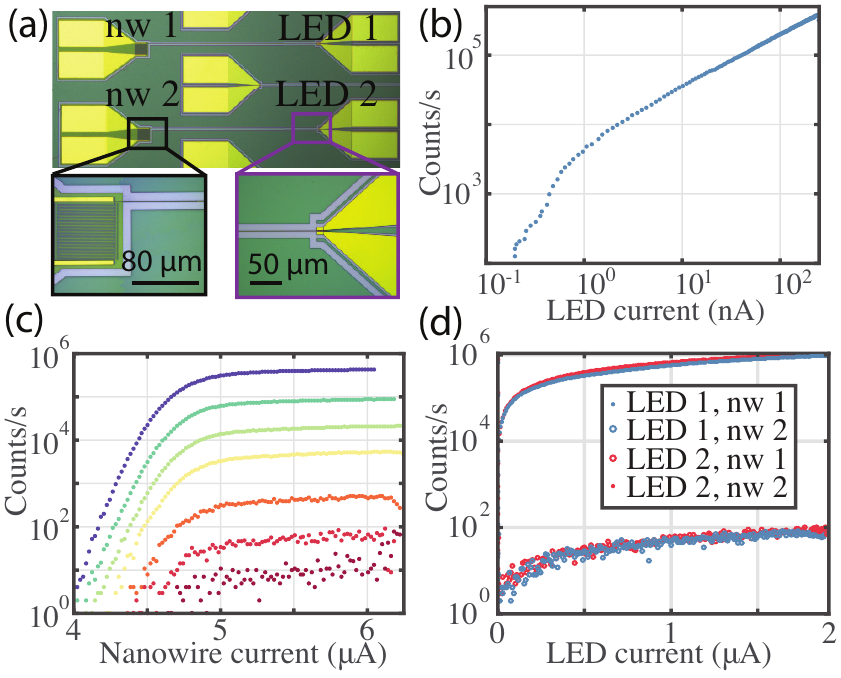}
 \caption{\label{fig:LED2nw}(a) Optical microscope image showing two LED to SNSPD (nw) structures. (b) Counts per second versus LED current for SNSPD bias current of 5 $\mu$A for one of the structures shown in part (a). (c) SNSPD counts per second versus nanowire bias current at different LED currents for one of the structures shown in part (a). LED currents of 100 pA,  150 pA, 320 pA, 1 nA, 4 nA, 28 nA, 250 nA are shown. (d)  SNSPD counts  per second versus bias current for the two different LEDs and SNSPDs indicated in part (a), indicating the emitted light is waveguide coupled.}%
 \end{figure}

The LED-to-nanowire devices were next tested in a sorption pump cryostat at 800 mK. An optical microscope image of these devices is shown in Fig. \ref{fig:LED2nw} (a). In Fig. \ref{fig:LED2nw} (b) the SNSPD bias current is fixed at 5 $\mu$A while the LED current bias is increased. We observe an increase in counts on the SNSPD above the background noise level when the LED bias current reaches 150 pA. In Fig. \ref{fig:LED2nw} (c) we show the SNSPD response as a function of current through the nanowire for seven values of LED bias current. For each LED current, there is an initial increase in counts per second as the SNSPD bias current is increased, and the response levels off at higher bias current as the internal quantum efficiency of the SNSPD saturates \cite{Marsili2013}. To verify that the light is waveguide-coupled, the crosstalk between two adjacent devices is examined, as shown in Fig. \ref{fig:LED2nw} (d).  SNSPD1 and LED1 are connected via a waveguide, and SNSPD2 and LED2 are connected via a separate waveguide. When the applied current to LED1 is scanned we observe 40 dB higher response on SNSPD1 than on SNSPD2. The reverse is true when the current is scanned on LED2. There is no increase in count rate when the LED is reverse biased at the same voltage level. The total system efficiency of 5 $\times 10^{-7}$ is calculated by multiplying the photon energy by the measured detector count rate and dividing the product by the total electrical power. This includes losses due to nonradiative recombination in the LED, light not coupled to the waveguide, light in the waveguide not absorbed by the SNSPD, and inefficiencies in the SNSPD.

\begin{figure}
\includegraphics[width=8.5cm]{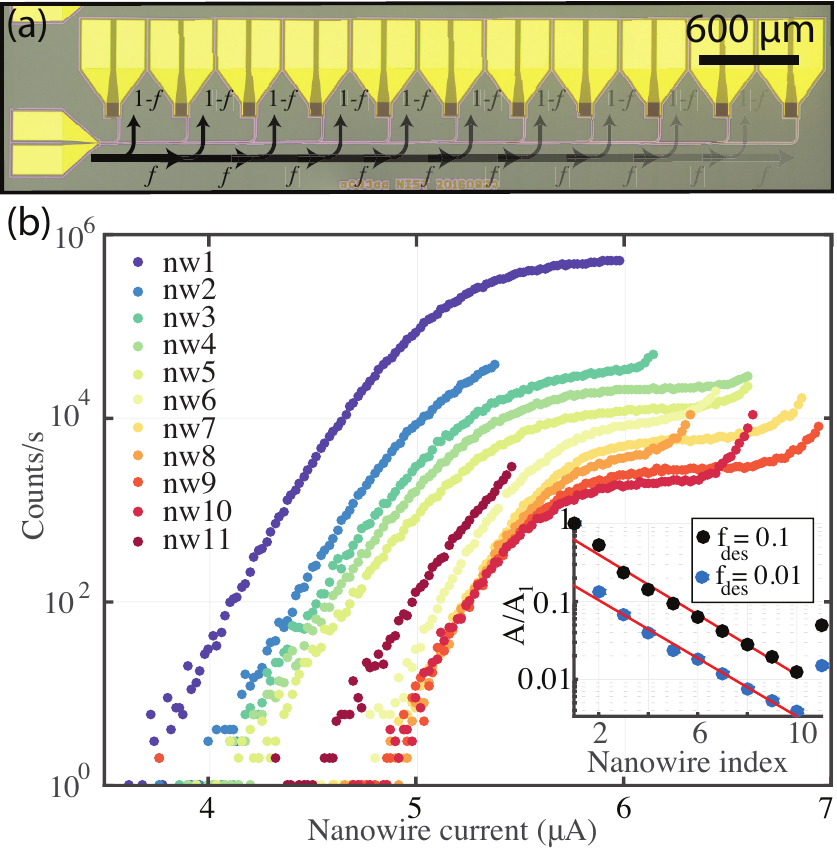}
 \caption{\label{fig:HiDRA}(a) Optical microscope image of the HiDRA, designed to split off a fraction of the light to each of the 11 detectors. (b) Counts versus nanowire current for all 11 detectors. Inset: Plot of detector response (defined in text) versus nanowire index for two different HiDRA structures designed to have different beamsplitting ratios. The red line is a linear fit from which the beamsplitting ratio is calculated.}%
 \end{figure}
 
To demonstrate the scalability of LEDs and SNSPDs fabricated with the process presented here, a device with an LED coupled to eleven SNSPDs was demonstrated. An optical microscope image of the device is shown in Fig. \ref{fig:HiDRA} (a). The device consists of an LED (on the left) that couples into a waveguide with a series of power taps. At each tap, a fraction, $f$, of the light is directed to subsequent detectors, while $1-f$ of the light goes to the nanowire detector at that port. This design allows measurement of a high dynamic range of light, where the $n^{th}$ detector will receive $(1-f)f^{n-1}$ of the light, assuming consistent power taps and detectors. For example, if the beamsplitting ratio is 1:9, $f$ = 0.1, and the first detector will receive 90$\%$ of the light while the tenth detector will receive $9\times10^{-8} \%$ of the light. In principle this should allow detection from the single photon level up to milliwatts of optical power, limited by the cryostat cooling power and the background scattered light level on-chip. 

This high-dynamic-range detector array (HiDRA) is useful for characterizing the operation of the LED over a broad range of current injection levels. Figure \ref{fig:HiDRA} (b) shows the counts versus nanowire bias for eleven SNSPD detectors comprising a HiDRA. Since the intensity of the light on each SNSPD for a fixed LED intensity $I_0$ can be determined from $I = I_0(1-f)f^{n-1}$, the beam splitting ratio $f$ can be extracted from the slope of a semilog plot of the SNSPD counts versus SNSPD index $n$. To generate such a plot, each of the curves shown in Fig. \ref{fig:HiDRA} (b) was fit to an error function, and the amplitude of each error function, $A$,  (normalized by the value for the first detector, $A_1$) was plotted versus the nanowire index. 

The results of this plot for two different HiDRAs, designed to have two different beamsplitting ratios, are shown in the inset to Fig. \ref{fig:HiDRA} (b). The fit was performed leaving out the first and last detectors, for reasons described below. These HiDRAs were designed to have $f_{des}$ = 0.1 and $f_{des}$ = 0.01. However, the device did not perform as designed and the beamsplitting ratio that we observed was significantly larger than this, with $f = 0.65 \pm 0.07$ for both, as calculated from the figure inset in Fig. \ref{fig:HiDRA} (b). This discrepancy is likely due to coupling of the LED to higher-order modes of the structure. The ridge waveguide with deep-etched sides (see diagram Fig. 2 (a)) supports a multitude of other slab modes, and light from the LED likely couples to these modes. 

This hypothesis is supported by the observed deviation of the first detector response from the expected straight line. In fact, for the designed $f_{des}$ = 0.1 ($f_{des} = 0.01$) structure, this point indicates that the splitting ratio between the first and second detector $f_{12}$ is 0.5 (0.15).  This may be due to the fact that there is a significant contribution from the fundamental mode which behaves according to design and is mostly directed to the first detector, with a larger proportion sent to the first detector for the $f_{des}$ = 0.01 detector. At subsequent detectors the contribution of the fundamental mode is negligible and higher order modes dominate the response. 

Additionally, the final detector falls off the expected line. This is because the final detector has no beamtap. The expected value for the intensity is therefore $I_0(f^{11})$, which leads to the expected value of the final point to fall between the intensity values for $n$ = 7 and $n$ = 10. We observe the value to fall between $n$ = 6 and $n$ = 7, indicating $f \approx 0.76-0.78$. This is outside the range of expected values calculated from the slope, likely due to the multimode nature of the waveguide.
 
Thirty-six SNSPDs from this wafer were tested, and all were functional. As can be seen in Fig. \ref{fig:HiDRA} (b), the critical current and the length of the plateau region varied. This is likely due to the fact that the SNSPD widths were close to the resolution limit of the photolithography tool, and as a result suffered from variability. The variability was predictable from die to die for a particular pattern, indicating that it was caused by the different doses received at different parts of the chip. This could be accounted for by using a fill pattern to make the dose more uniform, or simply by using a higher resolution lithography tool. Twenty LEDs were tested, and fifteen were functional. We attribute the lower yield of the LEDs to incomplete removal of the masking resist after the implant stages.  
  
 \section{Discussion}
 
Waveguide-integrated single-photon detectors integrated with electrically injected light sources are promising for many applications. The devices demonstrated here were chosen as the first of a suite of on-chip characterization devices. Future devices will include on-chip spectrometers \cite{Pathak2014}, Hanbury Brown-Twiss \cite{Schuck2016} beamsplitters and slow-light waveguides or resonators for enhanced emission \cite{Lund-Hansen2008,Sumikura2014}. This platform could also be used as a testbed to determine the optical properties of other defects in Si \cite{Davies1987}. SNSPDs have been observed to work in the mid-IR \cite{Li2017}, which will allow characterization of deep mid-IR defects in Si. For example, chalcogenide defects have been proposed as a spin-photon qubit in Si \cite{Morse2016}. 

It may also be possible that the W-center itself exhibits single photon emission. By integrating an on-chip beamsplitter with single photon detectors and performing ion implantations with different densities of emitters, it will be possible to test whether or not this is the case \cite{Khasminskaya2016}. It is also possible that when coupled to a resonator, emissive centers in Si will provide sufficient gain for an on-chip laser which could then be used to generate pairs through cavity-enhanced four-wave mixing \cite{Gentry2015}. Optical gain and stimulated emission have already been demonstrated with other defects in Si \cite{Cloutier2005}.

These detectors and LEDs could be integrated into optoelectronic neurons, as proposed recently \cite{Shainline2017}. For these LEDs to enable the massive scaling in neuromorphic systems, the efficiency will need to be increased by several orders of magnitude. There are a variety of avenues that can be pursued for this improvement. 

The easiest is to improve the coupling to the SNSPD. If light were coupling only to the fundamental mode of the waveguide, finite-difference eigenmode simulations show that between 10$\%$ and 50$\%$ of the light would be absorbed by the 100 $\mu$m long nanowire detectors. This can be easily improved by increasing the detector length. However, some of the light is coupled to other modes of the waveguide, which is absorbed even less efficiently by the SNSPD. More careful design of the overlap of the optical mode with carriers during current injection may lead to significant improvements, both by increasing electrical injection efficiency, and the fraction of light coupled into the fundamental waveguide mode.  Therefore, we expect at least an order of magnitude improvement solely through improved waveguide and detector implementation, enabled by higher resolution lithography. 

Additionally, light is being lost from the back side of the LED. Putting a mirror behind the waveguide to reflect that light back, or simply using that light elsewhere on chip would provide additional gains. The optical intensity is also not proportional to the implanted defect density \cite{Harding2006}, and it is possible that a lower density of emitters will have fewer non-radiative recombination centers and will allow us to reach higher efficiencies, although at the cost of a lower maximum saturated intensity. 

Moving to a higher resistivity Si wafer with fewer non-radiative recombination centers may also increase the efficiencies, as a Si LED with $1\%$ efficiency has been realized by reducing the non-radiative pathways \cite{Green2001}. Additionally, B-doping has been shown to quench the W center emission \cite{Charnvanichborikarn2010}, so moving to an intrinsic wafer may have an added benefit in this case. The emission into the optical mode may also be enhanced by the Purcell effect in a slow light optical waveguide \cite{Lund-Hansen2008} or an optical resonator \cite{Sumikura2014,LoSavio2011}, which can routinely provide enhancements of a factor of 5. Finally, there are other emitters that could prove more efficient than the W-center. There may be deep defects in Si that emit in the mid-IR but have yet to be explored due to the lack of adequate detectors in this region. With integrated SNSPDs, barriers to exploring these options are reduced.

We have demonstrated waveguide integrated W-center LEDs in Si with WSi SNSPD detectors in an on-chip optical link. We have also demonstrated these elements in a more complex integrated structure. This scaling is enabled by the fact that the LEDs are simple to fabricate via photolithography and ion implantation, and amorphous WSi SNSPDs have high yield and are easily deposited on virtually any substrate. Integration of these devices with fully etched single mode Si waveguides will allow higher performance and more complex optical devices. The platform can be applied to any ion-implanted emissive centers in Si that can be electrically injected.\\

This is a contribution of NIST, an agency of the US government, not subject to copyright.

%


\bibliography{LowTemperatureSiLEDs}
\end{document}